\documentclass[%
 aip,
 amsmath,amssymb,
 reprint,%
]{revtex4-1}

\usepackage{graphicx}% Include figure files
\usepackage{dcolumn}% Align table columns on decimal point
\usepackage{bm}% bold math
\usepackage[utf8]{inputenc}
\usepackage[T1]{fontenc}
\usepackage{mathptmx}
\usepackage{etoolbox}

\makeatletter
\def\@email#1#2{%
 \endgroup
 \patchcmd{\titleblock@produce}
  {\frontmatter@RRAPformat}
  {\frontmatter@RRAPformat{\produce@RRAP{*#1\href{mailto:#2}{#2}}}\frontmatter@RRAPformat}
  {}{}
}%
\makeatother
\begin{document}

\preprint{AIP/123-QED}

\title{Intermediate-temperature specific heat of solids and the rationale behind the Maier-Kelley empirical formula}
%\title{First-principles high-temperature specific heat of solids and the rationale behind the Maier-Kelley empirical formula}
% Force line breaks with \\
\author{Valmir Ribeiro}
 \altaffiliation{Departamento de
		F\'{\i}sica, Centro de Ci\^encias Exatas e da Natureza, Universidade Federal de Pernambuco, Recife, Pernambuco
		50670-901 Brazil}
\author{Fernando Parisio}
\email{fernando.parisio@ufpe.br}
 \altaffiliation{Departamento de
		F\'{\i}sica, Centro de Ci\^encias Exatas e da Natureza, Universidade Federal de Pernambuco, Recife, Pernambuco
		50670-901 Brazil}%Lines break automatically or can be forced with \\

\date{\today}% It is always \today, today,

\begin{abstract}
The heat capacity of solids at intermediate-to-high temperatures  is of fundamental importance to several fields ranging from geology to material science. It depends on a variety of factors, with anharmonicity and, ultimately, melting playing a pivotal role.  In this work we develop a first-principles model from an analytically tractable semi-harmonic oscillator Hamiltonian. The resulting specific heat expression depends not only on the Einstein temperature of the material but also on other physical parameters. We compare our predictions with experimental data for copper, aluminum, lead, silicon, and germanium with rather satisfactory results, especially considering that {\it there are no fitting parameters} in our theory. We finish this work by showing that our results formally justify the otherwise purely empirical formula by Maier and Kelley, also providing its coefficients in terms of elementary physical quantities.
\end{abstract}

\maketitle

\section{Introduction}
The first attempts to understand the thermo-mechanical behavior of metals mark one of the starting points of solid-state physics
as an independent field of research. Even the purely classical prediction of Dulong and Petit plays a historical role that is often underestimated. It not only constitutes a predictive success of statistical mechanics but also played a central role in Mendeleev's classification of elements in what became the periodic table \cite{dp}.  

At a fundamental level, much attention is given to the low-temperature limit, due to the radical discrepancy between the classical and quantum predictions in this regime. However, understanding heat capacities in the intermediate-to-high temperature regime is of fundamental importance to a variety of applied fields, like metallurgy \cite{MK}, geology\cite{geology, geology2} and volcanology \cite{volc}, ceramics \cite{ceramics}, and materials science \cite{material}, in general.

There is a multitude of models aiming at a more detailed description of heat capacities of solids. Some examples are modified Einstein models \cite{CP-1,CP0}, merges of Debye and Einstein models \cite{ED0,ED1}, consideration of anharmonic potentials \cite{anh,anh2,anh3,anh4}, as well as closed analytical expression approximating the Debye integral \cite{CP1} and adaptations of the Debye model to encompass amorphous solids \cite{CP2}. We stress that the large majority  of models aiming at quantitatively precise descriptions employ multiple fitting parameters, see for instance \cite{vassilev}

It is, thus, important to justify the worthiness of a new framework. The model we introduce here is derived from first-principle considerations on a simple, analytically tractable oscillator, plus an input from quantum field theory. It, thus, sheds light on foundational aspects. Differently from the large majority of the models presented in the literature which try to reproduce experimentally observed deviations from the Dulong-Petit regime, the present model has no fitting parameters and, yet, yields a good agreement with experimental data, for several substances. In addition, it is able to unveil the justification on why the Maier-Kelley empirical ``expansion'' is efficient in describing the specific heat of  many solids at high temperatures. 

\section{The semi-harmonic oscillator}
The essential ingredient of elementary models describing  the specific heat of solids is a large number of localized, non-interacting oscillators (in this letter we will not address electronic properties). The most known examples are the Einstein and the Debye models \cite{callen}. In spite of the great success of these models to explain why $C_v$ vanishes as $T\rightarrow 0$ K, they are contrived in the sense that the Dulong-Petit regime persists for arbitrarily large temperatures. This is mainly due to the harmonic character of the considered oscillators. Thermal expansion, deviations from the Dulong-Petit plateau, and, ultimately, melting are intrinsically related to anharmonicity, which is typically taken into account by considering non-linear interactions. With this, however, the possibility to derive simple analytical results is hindered. 
 
Here we argue that a simple and effective way to take these aspects into account, in a closed analytical way, is to consider the quadratic Hamiltonian
\begin{equation}
\label{Ham}
{\cal H}=\frac{1}{2m}p^2+\frac{1}{2}m\omega_0^2 q^2-W qp,
\end{equation}
for an oscillator of mass $m$. The quantity $\omega_0$ corresponds to the system's angular frequency for $W=0$, since, in this limit, we obtain the standard harmonic oscillator. This somewhat mysterious Hamiltonian has a simple physical interpretation.
%One may initially think that the term $Wqp$ would smoothly introduce anharmonicity in the system as $W$ varies. However, this is not exactly the case. 
The solution of Hamilton's equations readily gives:
\begin{equation}
q(t)=q_0\cos \Omega t+\tilde{q}\sin \Omega t, \,  p(t)=p_0\cos \Omega t+\tilde{p}\sin \Omega t,
\end{equation}
where $q_0$ and $p_0$ are the initial conditions in phase space, $\tilde{q}=(p_0-mW q_0)/(m\Omega)$ and $\tilde{p}=(Wp_0-m\omega_0^2 q_0)/\Omega$, see the Supplementary Material. Finally, and most importantly, the angular frequency reads:
\begin{equation} 
\label{freq}
\Omega=\omega_0\sqrt{1-\left(\frac{W}{\omega_0}\right)^2}.
\end{equation}
\begin{figure}[h]
  \includegraphics[height=5cm]{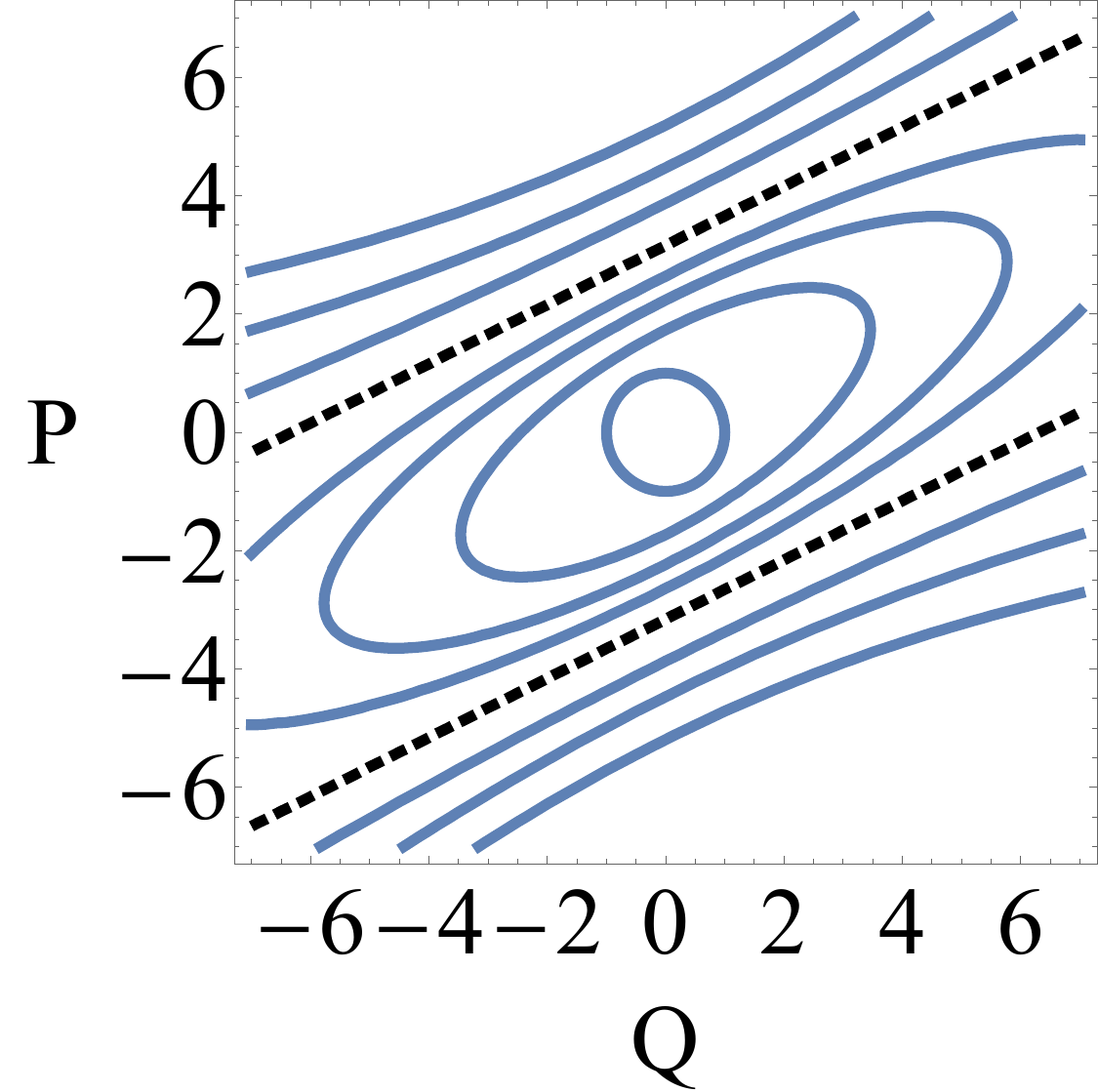}
  \caption{A diagram in phase-space for the classical system described by Hamiltonian (\ref{Ham}), for several values of $W$. For $W<\omega_0$, the dynamics is harmonic with the elipses becoming increasingly ``stretched'' as $W$ grows. For $W=\omega_0$ the motion becomes unbounded (dotted straight lines), while for $W>\omega_0$ the dynamics is hyperbolic. We employ the dimensionless variables $Q=q/b$ and $P=bp/\mathfrak{h}$, with $b=\sqrt{\mathfrak{h}/m\omega_0}$ and $\mathfrak{h}=1$ J.s . Note carefully that for any fixed value of $W\ne \omega_0$ the phase-space portrait is either completely composed by ellipses or completely composed by hyperbolas.}
  \label{fig1}
\end{figure}
We see that, while the motion described by $q(t)$ and $p(t)$ is described by linear combinations of sine and cosine functions, whenever $W<\omega_0$; it becomes hyperbolic, and thus unbounded, for $W>\omega_0$, with a linear separatrix for $W=\omega_0$. Therefore, the physical interpretation of a system described by Hamiltonian (\ref{Ham}) is that of a harmonic oscillator with a built-in rupture point, see Fig. \ref{fig1}. This rupture, the transition from bounded to unbounded motions, is potentially useful in modeling high-temperature specific heats, because it should be somehow connected with melting. Hereafter we will refer to this system as the semi-harmonic oscillator.

%A final comment on this system is that Hamiltonian (\ref{Ham}) cannot be brought to the standard form ${\cal H}={\cal P}^2/2+\Omega^2{\cal Q}^2/2$ through some real, linear canonical transformation. So, there is no way to associate the semi-harmonic oscillator with a conveniently transformed simple harmonic oscillator (the two systems are fundamentally inequivalent). For details on the classical solution, see the Supplementary Material.

We are mainly interested in the quantum counterpart of the semi-harmonic oscillator. The symmetrized canonical quantization of (\ref{Ham}) reads $\hat{\cal H}=\frac{1}{2m}\hat{p}^2+\frac{1}{2}m\omega_0^2 \hat{q}^2-\frac{W}{2} (\hat{q}\hat{p}+\hat{p}\hat{q})$.
This Hamiltonian can be expressed in terms of standard ladder operators $\hat{a}=(\hat{q}/b+ib\hat{p}/\hbar)/\sqrt{2}$ and $\hat{a}^{\dagger}=(\hat{q}/b-ib\hat{p}/\hbar)/\sqrt{2}$, with $b=\sqrt{\hbar/(m \omega_0)}$, resulting in a cumbersome expression (see the Supplementary Material). However, there exists a Bogoliubov transformation such that 
\begin{equation}
\hat{\cal H}=\hbar \Omega(\hat{c}^{\dagger}\hat{c}+1/2), 
\end{equation}
with $[\hat{c},\hat{c}^{\dagger}]=1$ and $\Omega$ given by (\ref{freq}). Note that $\hat{\cal H}$ is Hermitian only if $W\le\omega_0$. Whenever this condition is fulfilled, the new bosonic operators are given by 
\begin{equation}
\hat{c}=\sqrt{\frac{\omega_0}{2\Omega}}\left(\frac{\hat{q}}{b}+ie^{i\phi}\frac{b\hat{p}}{\hbar}\right),\; \hat{c}^{\dagger}=\sqrt{\frac{\omega_0}{2\Omega}}\left(\frac{\hat{q}}{b}-ie^{-i\phi}\frac{b\hat{p}}{\hbar}\right),
\end{equation}
with $\cos \phi=\Omega/\omega_0$. The excitations associated with these ladder operators belong to the class of ``softened'' phonons, as we will see next. 
Therefore, with the previous elementary approach one obtains genuine, non-interacting quasi-particles, provided $W<\omega_0$.

\section{Statistical considerations: a closed formula for $C_p$.}
Now, we consider an ensemble of semi-harmonic oscillators in contact with a heat bath at absolute temperature $T$. A natural way to bring a flavor of anharmonicity to the model is to assume that the rupture occurs as $T$ reaches a finite temperature $T^*$, i. e., that the Hamiltonian (\ref{Ham}) is temperature dependent through $W=W(T)$. The use of temperature-dependent effective Hamiltonians in modeling condensed-mater systems is usual, see for instance \cite{pina,gudyma, horbach}. The once localized oscillators would become unbounded particles at $T^*$, satisfying $W(T^*)=\omega_0$. With $\Omega$ depending on $T$ we have both a rupture temperature and anharmonicity (through a temperature-dependent frequency), which would correspond to a continuous softening of phonons, in an analytically tractable way. 

At first sight, one might consider that $T^*$ should be directly identified with the melting temperature $T_m$, but this is not the case. The reason is clear from the analysis of the classical system: whenever we have $W\rightarrow \omega_0$ ($T\rightarrow T^*$), the amplitudes of $q(t)$ and $p(t)$ diverge (see the expressions for $\tilde{q}$ and $\tilde{p}$), that is, one of the axes of the elipses in Fig. 1 becomes arbitrarily large as $T\rightarrow T^*$. Of course, actual melting happens much before this point (divergence of the oscillator's amplitude) and, thus, $T^*$ should be viewed as an {\it upper bound} for the actual melting temperature. We will see that this is indeed the case in dealing with several substances to be considered later.

In addition, the functional form of equation (\ref{freq}) is quite suggestive if one considers that a more advanced treatment of anharmonic effects, at the level of quantum field theory (QFT), leads to the following temperature-dependent angular frequency: $\Omega=\omega_0\sqrt{1-6\alpha \gamma T}$, where $\alpha$ is the linear thermal expansion coefficient (i. e., $\alpha_{\rm vol}\approx 3\alpha$) and $\gamma$ is the Gruneisen parameter of the material \cite{callen,daniel}. This expression is directly compatible with Eq. (\ref{freq}) with the simple association $W^2(T)/\omega_0^2=6\alpha \gamma T$, and, therefore
\begin{equation}
\label{rup}
%T^*=\frac{1}{6\alpha \gamma}.
T^*=(6\alpha \gamma)^{-1}.
\end{equation}
Note, however, that the QFT result comes from complicated perturbative calculations over anharmonic potentials, which do not lead to simple, closed analytical results for the specific heat, in general. The correspondence in (\ref{rup}) indeed leads to $T^*>T_m$. For a large variety of solids the Gruneisen parameter is such that $1<\gamma<4$, while for the thermal expansion coefficient we have $10^{-4} {\rm K}^{-1}<\alpha <10^{-6}{\rm K}^{-1}$, thus, we have $10^4 {\rm K}<T^*<10^6 {\rm K}$, roughly. Since the highest melting temperatures, at normal pressure, are around $4000$ K,
we have $T_m/T^*< 1$. For the three metals we address in what follows we found $2.39<T^*/T_m<2.91$ for metals, and $T^*/T_m\approx 44$ and $22$, for Si and Ge, respectively .

We must be careful in dealing with thermodynamic relations involving temperature-dependent Hamiltonians. In the canonical formalism, a variation in the reservoir temperature leads to a repopulation of the energy levels according to the Maxwell-Boltzmann statistics, for sufficiently high temperatures. On the other hand, if the energy levels themselves depend on $T$, a change in temperature would produce two effects: a shift in the energy levels and a repopulation. In a quasi-static processes 
equilibration takes place after each infinitesimal temperature change and it still holds that the average energy after each step is given by
\begin{equation}
U\equiv \langle \hat{\cal H} \rangle= \frac{1}{\cal Z}\sum_nE_n(T)e^{-\frac{E_n(T)}{k_BT}},
\end{equation}
where ${\cal Z}$ is the canonical partition function and $k_B$ is the Boltzmann constant. Note, however, that $\langle \hat{\cal H} \rangle \ne -(d/d\beta)\ln {\cal Z}$, with $\beta^{-1}=k_BT$.

The molar specific heat (at constant volume) is given by $C_v=3N_A dU/dT$, with $E_n=\hbar\Omega(T)(n+1/2)$ and $N_A$ being Avogadro's number. The majority of experiments are done under constant pressure and, thus, it is more convenient to deal with the isobaric molar specific heat, $C_p=(1+3\alpha\gamma T)C_v$, which, after straightforward manipulations, reads
%
%\begin{eqnarray}
%\nonumber
%C_v&=&3R\left( \frac{T_E}{2T}\right)^2(1-3\alpha\gamma T) \left[\sinh\left(\frac{T_E}{2T}\sqrt{1-6\alpha \gamma T} \right) \right]^{-2}\\
%\label{cv}
%&-&3R\frac{3 \alpha \gamma T_E}{2\sqrt{1-6\alpha \gamma T}} \coth\left(\frac{T_E}{2T}\sqrt{1-6\alpha \gamma T} \right),
%\end{eqnarray}
%
%
\begin{eqnarray}
\nonumber
C_p&=&3R\left( \frac{T_E}{2T}\right)^2[1-9(\alpha\gamma T)^2] \left[\sinh\left(\frac{T_E\sqrt{1-6\alpha \gamma T}}{2T} \right) \right]^{-2}\\
\label{cp}
&-&\frac{9R \alpha \gamma T_E(1+3\alpha\gamma T)}{2\sqrt{1-6\alpha \gamma T}} \coth\left(\frac{T_E\sqrt{1-6\alpha \gamma T}}{2T} \right),
\end{eqnarray}
where $T_E=\hbar \omega_0/k_B$ is the Einstein temperature of the material and $R=N_Ak_B\approx8.3145$ J mol$^{-1}$ K$^{-1}$. The previous equation is one of our main results and is valid for temperatures sufficiently high so that the Bose-Einstein and Maxwell-Boltzmann statistics do not differ appreciably. This is because  $\Omega=\omega_0\sqrt{1-6\alpha \gamma T}$ is an approximation of the more general expression $\omega(T)=\sqrt{\omega_0^2- c \,n(T)}$, where $n(T)$ is the average number of excited phonons and $c$ an appropriate constant factor \cite{daniel}. Since we are interested in intermediate to high temperatures, we will avoid this unnecessary complication. A safe lower bound is given by $T>T_E/2$. Notice that the second term in (\ref{cp}) never diverges since $T<T_m<T^*$. Note also that for $\alpha=0$, we recover the Einstein model.

A final, important remark is that the quantities $\gamma$ and, especially $\alpha$, in fact depend on the temperature $T$. Therefore, we resort to experimental sources that determine $\gamma$ and $\alpha$ for high temperature ranges, as compared to room temperatures. In order to standardize our procedure for the different substances addressed in what follows, we employ the measured values of $\gamma$ and $\alpha$ for the closest temperature to $T_m$.
\section{Comparison with experimental data}
We are now in a position to compare our theoretical predictions with experimental data. Here we address five pure, crystalline substances, namely, copper, aluminum, lead, silicon, and germanium. The choices are related to the abundance of experimental data on these elements in a wide range of temperatures and on their relevance in several technological applications. The source from which most values of $\alpha(T)$ have been obtained is \cite{alpha}, while data on the Gruneisen is more scarce and scattered. 
\begin{figure}[h]
    \includegraphics[height=4cm]{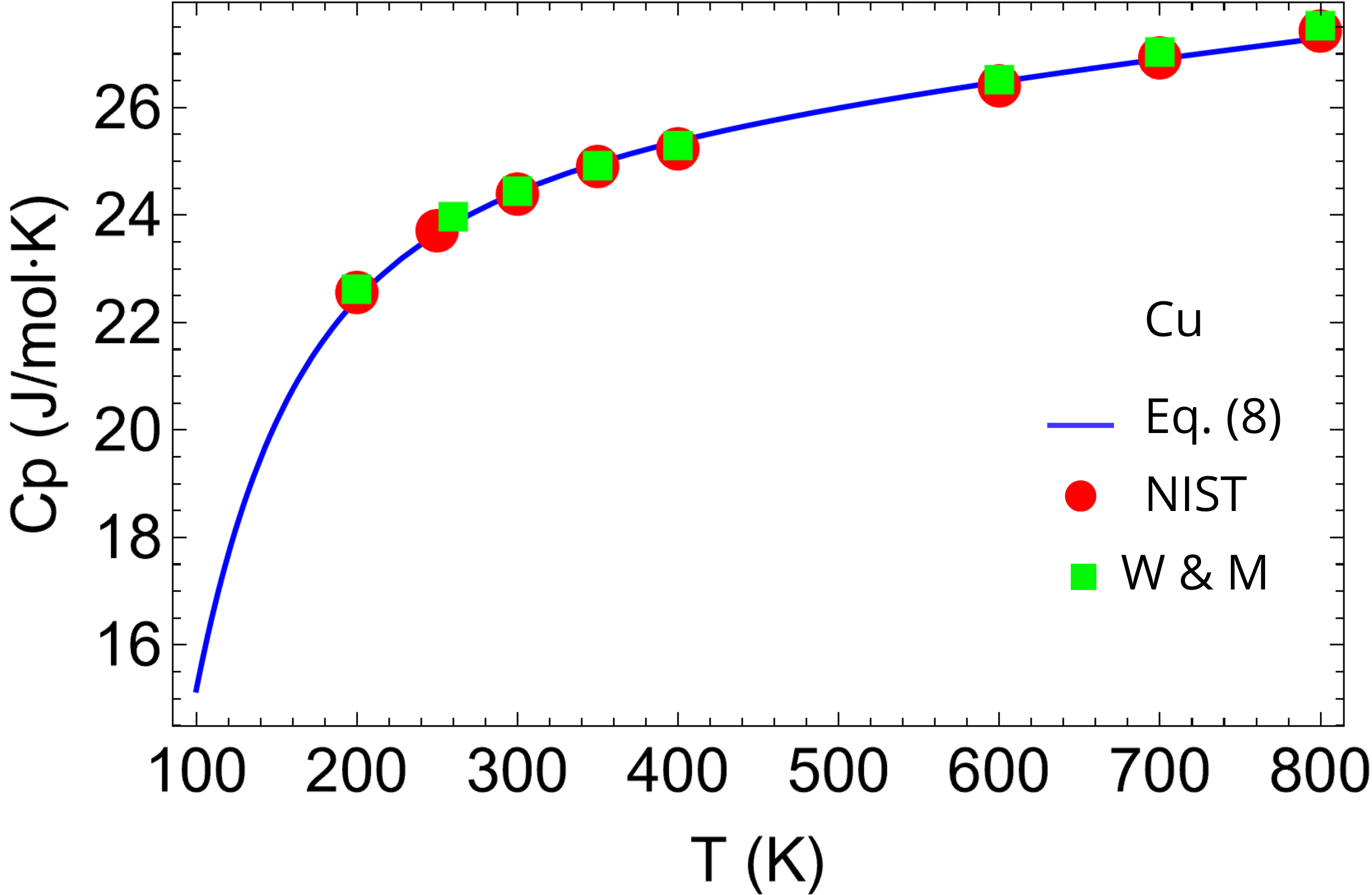}
  \caption{Copper: the blue curve represents Eq. (8), red bullets and green squares are experimental data from \cite{nistCu}, and reference \cite{WM}, respectively (no fitting).}
  \label{fig2}
\end{figure}
\begin{figure}[h]
  \includegraphics[height=4cm]{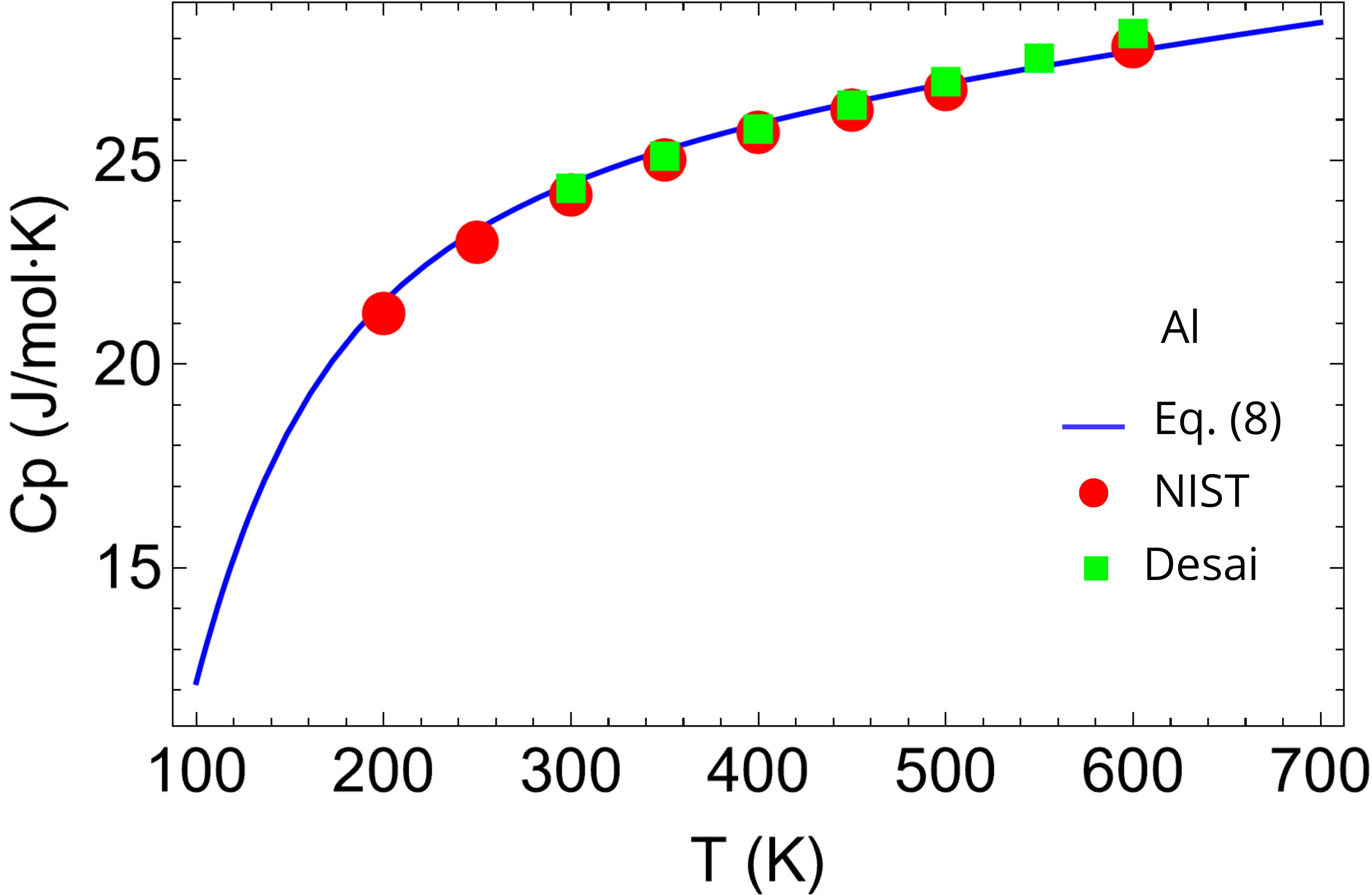}
  \caption{Aluminum: the blue curve represents Eq. (8), red bullets and green squares are experimental data from \cite{nistAl} and reference \cite{desai}, respectively (no fitting).}
  \label{fig3}
\end{figure}
\begin{figure}[ht]
  \includegraphics[height=4cm]{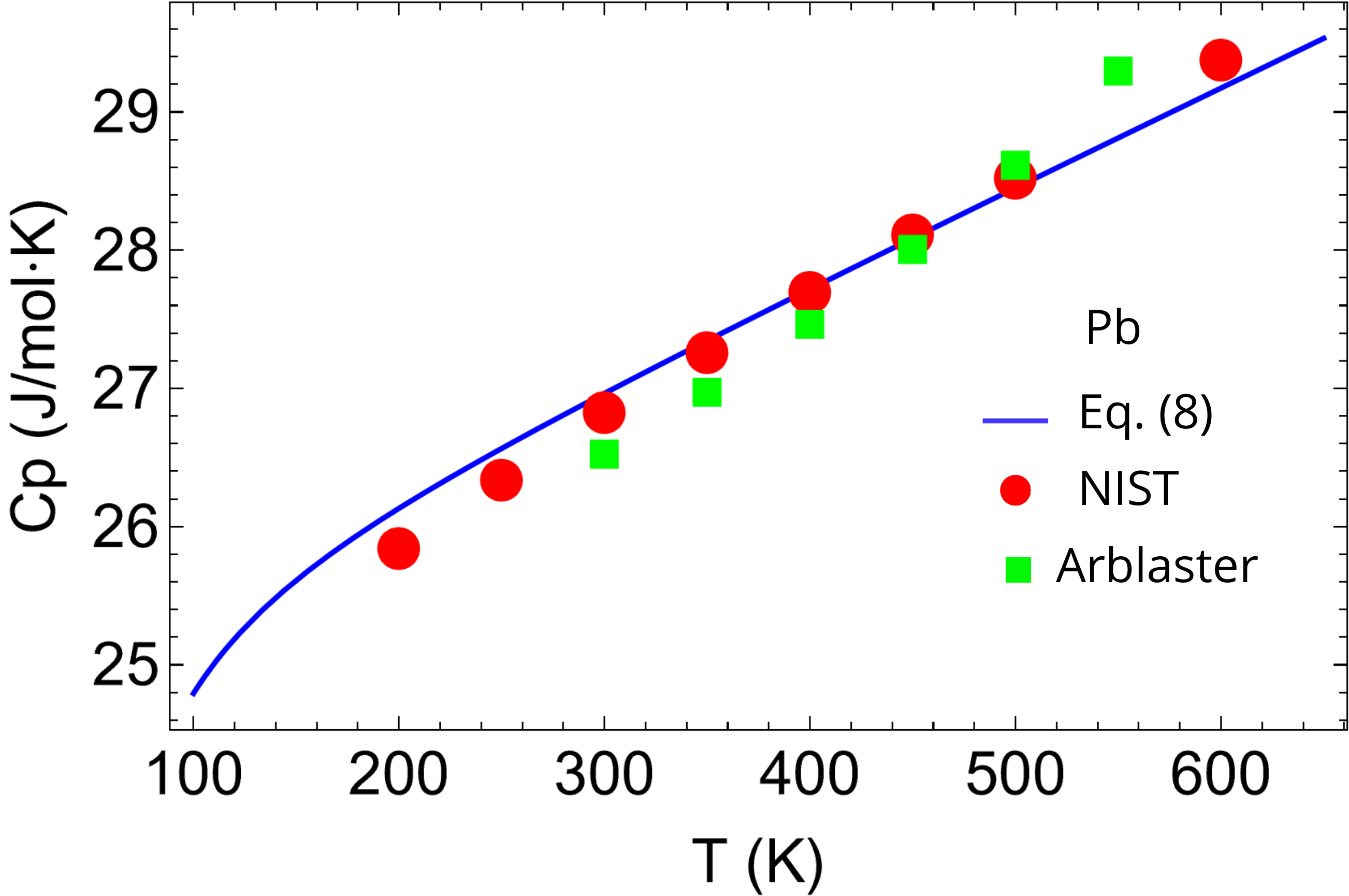}
  \caption{Lead: the blue curve represents Eq. (8), red bullets and green squares are experimental data from \cite{nistPb} and reference \cite{arblaster}, respectively (no fitting).}
  \label{fig4}
\end{figure}
\begin{figure}[h]
  \includegraphics[height=4cm]{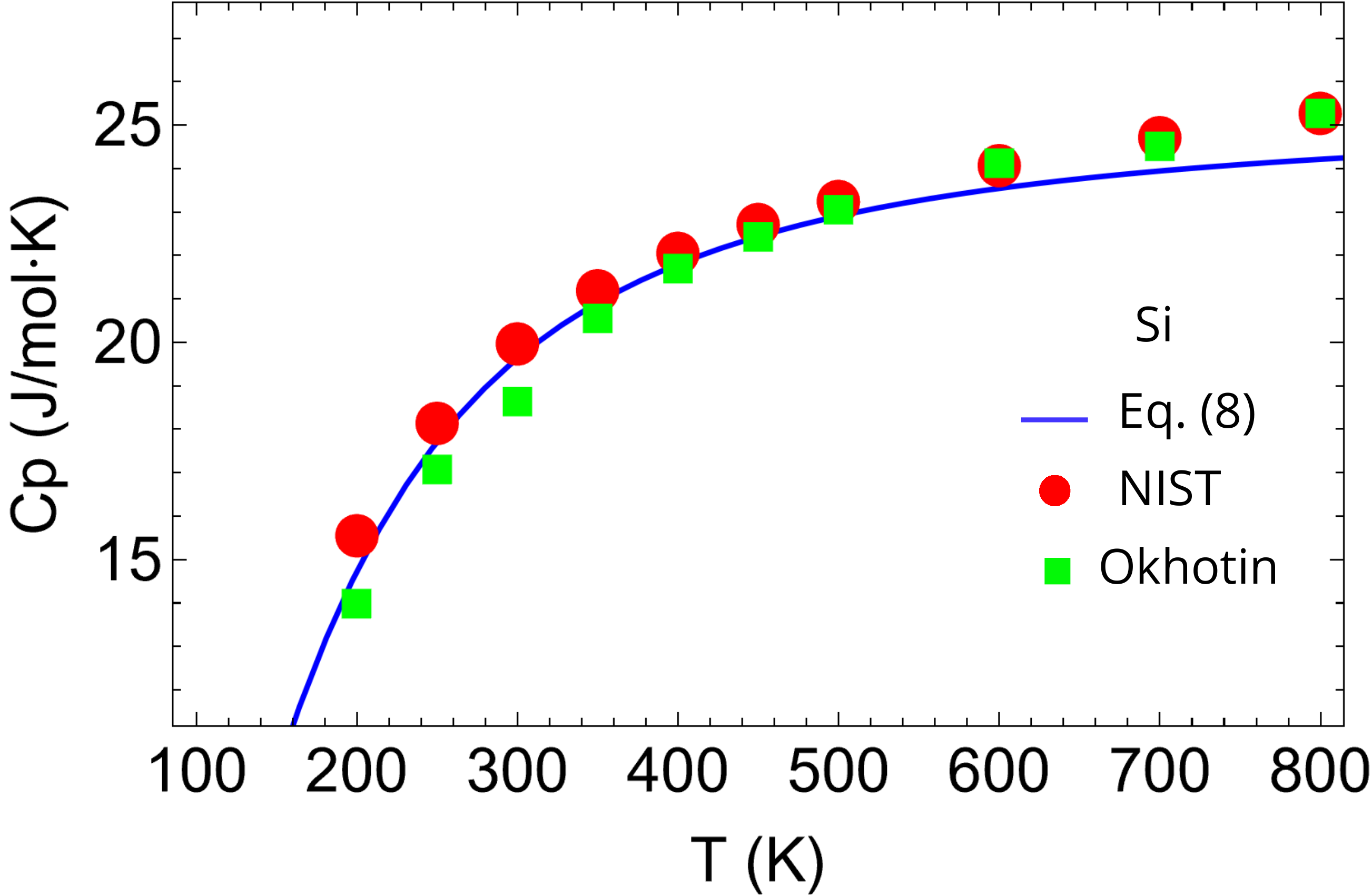}
  \caption{Silicon: the blue curve represents Eq. (8), red bullets and green squares are experimental data from \cite{nistSi} and reference \cite{okhotin}, respectively (no fitting).}
  \label{fig5}
\end{figure}
\begin{figure}[h]
  \includegraphics[height=4cm]{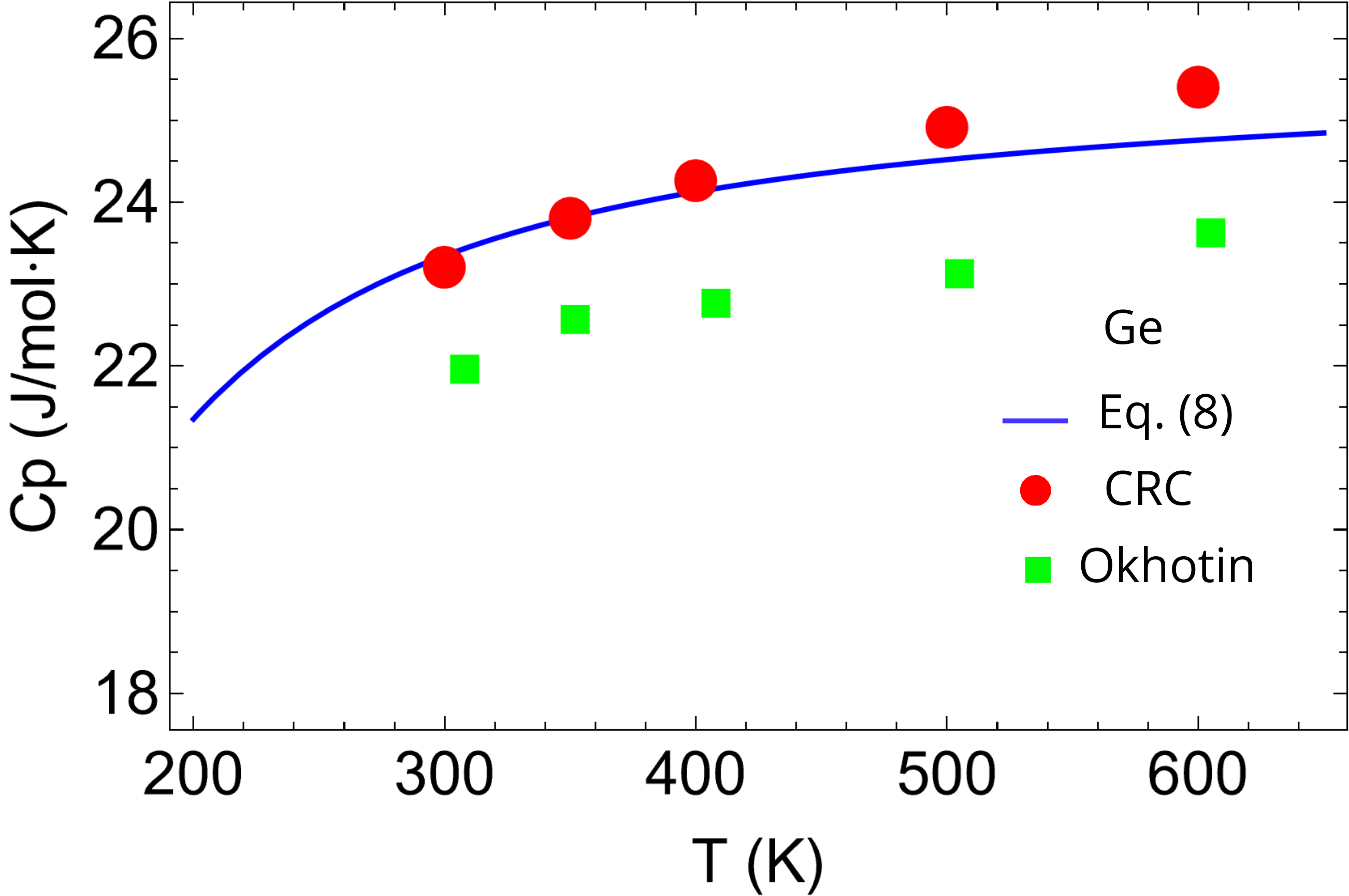}
  \caption{Germanium: the blue curve represents Eq. (8), red bullets and green squares are experimental data from \cite{nistGe} and reference \cite{okhotin}, respectively (no fitting).}
  \label{fig6}
\end{figure}

We start by investigating copper (Cu), a transition metal with $T_E\approx 248$ K \cite{hassel}. From \cite{alpha} we get $\alpha\approx 21.3 \times 10^{-6}$ K$^{-1}$, near $T=T_m$, and the Gruneisen parameter reads $\gamma\approx 2. 03$ at $T=400$ K, from \cite{chinese}. This leads to $T^* \approx 3.854$ K (compare with $T_m\approx 1358$ K).  By plugging these parameters in Eq. (\ref{cp}), we obtain the result shown as the continuous blue line in Fig. \ref{fig2}. For comparison, we also plot experimental data available at the NIST webpage \cite{nistCu} and from White and Minges \cite{WM}.

We now address the post-transition metals aluminum and lead. For pure Al, we have $T_E\approx 303$ K and, from reference \cite{alpha}, 
we obtain $\alpha=35.5 \times10^{-6}$ K$^{-1}$, in the vicinity of $T=T_m$.  from\cite{alpha}. The Gruneisen parameter of Al around this temperature is $\gamma \approx 2.1$ \cite{brooks}. Therefore, we get $T^*\approx 2.235$ K, while the melting temperature under a pressure of 1 ATM is $T_m\approx 933$ K.  The result is shown in Fig. \ref{fig3} together with experimental results form \cite{nistAl} (red bulets) and \cite{desai} (green squares). For pure Pb the employed parameters are $T_E\approx 63.6$ K \cite{khvan}, $\gamma\approx 2.6$ \cite{young}, and $\alpha\approx 36.6 \times 10^{-6}$ K$^{-1}$ , from \cite{alpha}. In Fig. \ref{fig4} we compare our result with two sets of experimental data from NIST \cite{nistPb} and from reference \cite{arblaster}. For lead we have $T_m\approx 601$ K and $T^*\approx 1751$ K.

Next, we address the metalloids silicon and germanium. For pure Si we have the following parameters: $T_E\approx 516$ K, $\alpha\approx 4.5 \times 10^{-6}$ K$^{-1}$ from \cite{alpha}, and $\gamma\approx 0.5$ at temperatures higher than 500 K \cite{wei, kagaya}. These parameters lead to $T^*\approx 74.000 $ K, much higher than the melting temperature $T_m\approx 1687$ K.
The comparison between theory and experiment is depicted in Fig. \ref{fig5}, with experimental data from NIST \cite{nistSi} (red bullets) and from \cite{okhotin}. For crystalline germanium we have $T_E\approx 278$ K, $\alpha\approx 8.0 \times 10^{-6}$ K$^{-1}$ from \cite{alpha}, and $\gamma\approx 0.76$, from \cite{hassel,gauster}. The two sets of experimental data, from \cite{nistGe} and from Okhotin \cite{okhotin}, present a sizable discrepancy, with the theoretical prediction closer to the data from  \cite{nistGe}. For germanium we have $T_m\approx 1211$ K, while $T^*\approx 27.000$ K.

Finally, we also tested our theoretical predictions for silver (Ag), with no agreement, even qualitative, between Eq. (\ref{cp}) and the experimental data. This illustrates the fact Eq. (\ref{cp}) is an approximation which does not take into account several aspects, like electronic contribution and crystalline structure, for instance. For several substances, to disregard these and other factors, makes Eq. (\ref{cp}) insufficient.

\section{Maier-Kelley formula explained}
Back in 1932, in a brief paper on how to efficiently fit thermodynamical data at high temperatures \cite{MK}, Maier and Kelley remark that: {\it ``It is apparent that at high temperatures specific heats of normally behaving substances in the solid or liquid state increase nearly linearly with the temperature, and that a subtractive term is needed which will be effective at temperatures lower than that where the equipartition value of specific heat is reached.''} Inspired by this observation they propose the following formula: 
$C_p\approx 3R \left(A+B\,T-C\,T^{-2}\right)$, where $A$, $B$, and $C$ are positive constants to be adjusted according to experimental data. It is then reported that measured specific heats of solids, at high temperatures, could be satisfactorily fitted by this expression, with a better performance than pure polynomial fittings (with more than three terms). 

Since then, the Maier-Kelley formula has been employed hundreds of times in a variety of applied fields, see \cite{yuan, cupid, denisova0} for recent illustrations. Despite this, it remains as a purely empirical expression with no fundamental physical justification. We provide such a justification in what follows. 

First, we call attention to a subtle point. There are two independent quantities which are small in (\ref{cp}), in the temperature window $T_E/2<T<T_m$, namely, 
\begin{equation}
\label{small}
\epsilon \equiv T_E/2T\;\; \mbox{and}\;\; \delta \equiv6\alpha \gamma T=T/T^*,
\end{equation} 
despite the existence of a single variable ($T$). The parameters $\epsilon$ and $\delta$ may vary independently from one substance to another, in terms of which the specific heat reads
\begin{eqnarray}
\nonumber
C_p&=&3R\left\{\epsilon^2[1-(\delta/2)^2] \left[\sinh\left(\epsilon\sqrt{1-\delta} \right) \right]^{-2}\right.\\
\label{cp2}
&-&\left.\frac{ \epsilon \delta(1+\delta/2)}{2\sqrt{1-\delta}} \coth\left(\epsilon\sqrt{1-\delta} \right)\right\}.
\end{eqnarray}
Therefore, we proceed an expansion in terms of both parameters up to second order.
Note that the argument of the hyperbolic functions is smaller than $1$ and their Laurent series $(\sinh x)^{-2}= x^{-2}-1/3+x^2/15+\cdots$ and $\coth x=x^{-1}+x/3-x^3/45+\cdots$, can be employed. By collecting terms up to second order ($\sim \epsilon^2, \delta^2, \epsilon \delta$), we get $C_p\approx 3R(1+\delta/2-\delta^2/4-\epsilon^2/3)$. Note that there is no first-order term in the variable $\epsilon$, anticipating the absence of a term proportional to $1/T$ in the Maier-Kelley formula. By using (\ref{small}) to obtain the explicit temperature dependence, we get
\begin{eqnarray}
\label{MK2}
\nonumber
C_p&\approx& 3R \left(A+B\, T-C\,T^{-2}-B^2 T^2\right)  ,\\
A&=&1, \; B=\frac{1}{2T^*},  \;\;\mbox{and}\;\; C= \frac{T_E^2}{12}.
\end{eqnarray}
The first three terms are exactly the Maier-Kelley expression, with the constant factors determined in terms of $T_E$ and $T^*$, while the last term ($\sim T^2$) is a well-known empirical correction \cite{robie,mostafa} (for a recent reference see \cite{denisova}), whose coefficient we find to be $-B^2$ ($B$ is the coefficient of the term $\sim T$). 
  
We remark that, for composite substances with $n$ atoms per molecule, the factor $3R$ must be replaced with $3nR$, in equations (\ref{cp}), (\ref{cp2}), and (\ref{MK2}).
For the sake of comparison, it is instructive to express result (\ref{MK2}) in terms of the dimensionless temperature $\theta=T/T_D$, where $T_D$ is the Debye temperature of the material, as, for instance in \cite{matthias}. From (\ref{MK2}) we get 
\begin{equation}
C_p\approx 3nR \left[ 1+\frac{\theta}{2\theta^*}-\frac{1}{12}\left(\frac{\pi}{6}\right)^{2/3}\theta^{-2}-\frac{\theta^2}{4{\theta^*}^2}\right],
\end{equation}
where we used the relation $T_E=(\pi/6)^{1/3}T_D$ (which is only an approximation when experimental values are considered) and $\theta^*=T^*/T_D$. 
It is interesting to note that the coefficient of $\theta^{-2}$ is a material-independent constant, whose absolute value is $\left(\frac{\pi}{6}\right)^{2/3}/12\approx 1/18.47\approx 0,054$. In equation (5) of \cite{matthias} the value of the same quantity is empirically set to $1/20=0.05$, independently of the material, which is in line with our first-principle result.

\section{Final Remarks}

In summary, we developed a heat capacity model from a simple, analytically tractable oscillator, which is capable of describing the deviations from the Dulong-Petit regime. Expression (\ref{cp}) also provides a basis from which the Maier-Kelley empirical formula can be formally derived. An interesting perspective for future work is to assume that $T_E$ and $T^*$ are free parameters and optimizing them for a given substance in a given temperature interval. Also, a more comprehensive application of Eq. (\ref{cp}) to a larger collection of solids, including composite substances, would be an interesting test for the model proposed here.
\begin{acknowledgments}
		The authors thank Reinaldo de Melo e Souza (UFF) for discussions on a preliminary stage of this research. This work received financial support from the Brazilian agencies Coordena\c{c}\~ao de Aperfei\c{c}oamento de Pessoal de N\'{\i}vel Superior (CAPES), Funda\c{c}\~ao de Amparo \`a Pesquisa do Estado de S\~ao Paulo (FAPESP - Grant 2021/06535-0), and Funda\c{c}\~ao de Amparo \`a Ci\^encia e Tecnologia do Estado de Pernambuco (FACEPE - Grant BPP-0037-1.05/24).
	\end{acknowledgments}
\appendix

\section{Semi-harmonic oscillator: classical description}

Here we give details of the classical-mechanical description of the semi-harmonic oscillator, whose Hamiltonian reads:
\begin{equation}
\label{hamiltonClass}
    \mathcal{H} = \frac{1}{2 m}p^2 + \frac{1}{2}m \omega_0^2 q^2 - W  q p,
\end{equation}
thus, being a conserved quantity: $\partial {\cal H}/\partial t=d {\cal H}/d t=0 \Rightarrow {\cal H}={\cal H}_0$.
Hamilton's equations give:
\begin{equation}
   \begin{split}
       \dot{p} =& -m \omega_0^2 q + W p,\;\;   \dot{q} =\frac{p}{m} - W q,
   \end{split}
\end{equation}
which can be decoupled by taking the second derivative with respect to time: 
\begin{align}\label{EDM}
    \ddot{q} = \frac{\dot{p}}{m} - W \dot{q} \Rightarrow 
     \ddot{q} = -\left(\omega_0^2 - W^2 \right) q = -\Omega^2 q,
\end{align}
where 
\begin{equation}
\Omega \equiv \omega_0 \sqrt{1 - \left(\frac{W}{\omega_0}\right)^2}
\end{equation}
 is the frequency that characterizes the harmonic motion of the system, whenever $W<\omega_0$. By setting $p(0)=p_0$ e $q(0)=q_0$, we can write the general solution as:
\begin{equation}
    \begin{split}
        q(t) &= q_0 \cos \Omega t + \tilde{q} \sin \Omega t, \\
p(t) &= p_0 \cos \Omega t + \tilde{p} \sin \Omega t,
    \end{split}
\end{equation}
with
 \begin{equation}
 \label{IC}
 \tilde{q}=\frac{p_0-mW q_0}{m\Omega},\;\;\tilde{p}=\frac{Wp_0-m\omega_0^2 q_0}{\Omega}.
 \end{equation}
Note, in particular, that $ \tilde{q}$ and $ \tilde{p}$ diverge as $W\rightarrow \omega_0$ ($\Omega \rightarrow 0$).
If, on the other hand, we set $W>\omega_0$, the above expressions become
\begin{equation}
    \begin{split}
        q(t) &= q_0 \cosh |\Omega| t + |\tilde{q}| \sinh |\Omega| t, \\
p(t) &= p_0 \cosh |\Omega| t + |\tilde{p}| \sinh |\Omega| t,
    \end{split}
\end{equation}
Therefore, typically, we either have harmonic, bounded motion ($W<\omega_0$) or hyperbolic, unbounded motion ($W>\omega_0$). For $\Omega=0$ ($W=\omega_0$) we have a linear behavior. In this case the conserved Hamiltonian becomes a perfect square: $(p/\sqrt{2m}-\sqrt{m/2}\omega_0q)=\pm{\cal H}_0$, or more explicitly
\begin{equation}
   p(t)=m\omega_0q(t)\pm\sqrt{2m{\cal H}_0},
\end{equation}
which correspond to the two linear separatrices mentioned in the main text.

%One may think that an appropriate canonical transformation, ${\cal Q}=Aq+Bp$ and ${\cal P}=Cq+Dp$, with real-valued $A$, $B$, $C$, and $D$, could bring the hamiltonian to the standard form ${\cal H}={\cal P}^2/2+\Omega^2{\cal Q}^2/2$. This is not true, because the requirement on the Poisson bracket, $\{{\cal Q},{\cal P}\}_{q,p}=1$, leads to $A=\sqrt{W^2}B\pm \sqrt{-\Omega^2B^2}$, which is imaginary whenever $\Omega$ is real-valued.

Finally, it is worth mentioning that, for $\omega_0>W$, the average energy in the canonical ensamble is unchanged as compared to that of the simple harmonic oscillator (HO). In this case, the partition function is
\begin{equation}
{\cal Z}=\int dq dp\; e^{-\beta {\cal H}(q,p)}=\frac{2\pi}{\omega_0\Omega\beta},
\end{equation}
with $1/\beta=k_BT$, while the average energy reads
\begin{equation}
U=\int dq dp\; {\cal H}(q,p)e^{-\beta {\cal H}(q,p)}=\frac{2\pi}{\omega_0\Omega\beta^2 {\cal Z}},
\end{equation}
which gives
\begin{equation}
U=k_BT,
\end{equation}
which is the same as the simple HO's average energy. In particular, we obtain the Dulong-Petit result for the specific heat per oscillator's degree of freedom.

\section{Semi-harmonic oscillator: quantum description}
Due to the non-commutativity of $\hat{q}$ e $\hat{p}$, the symmetrized canonical quantization of the system leads to:
\begin{equation}\label{Hquant}
    \hat{\cal H}=\frac{1}{2m}\hat{p}^2+\frac{1}{2}m\omega_0^2 \hat{q}^2-\frac{W}{2} (\hat{q}\hat{p}+\hat{p}\hat{q})
\end{equation}
Therefore, rewriting (\ref{Hquant}) with the standard ladder operators  \[
\hat{a} = \frac{1}{\sqrt{2}}\left( \frac{\hat{q}}{b} + i \frac{b \hat{p}}{\hbar} \right) \quad \text{and} \quad \hat{a}^\dagger = \frac{1}{\sqrt{2}}\left( \frac{\hat{q}}{b} - i \frac{b \hat{p}}{\hbar} \right),
\]
with 
$$b = \sqrt{\frac{\hbar}{m \omega_0}},$$ we obtain:
\begin{equation}\label{Haa}
    \hat{\cal H} = \frac{\hbar \omega_0}{2} \left(\hat{a}^{\dagger}\hat{a} + 1/2 \right) - \frac{i \hbar W}{2} \left({\hat{a}^{\dagger^2}} - \hat{a}^2 \right).
\end{equation}
Nevertheless, we can introduce new operators $\hat{c}$ and $\hat{c}^\dagger$ defined as linear combinations of $\hat{a}$ and $\hat{a}^\dagger$, namely, $\hat{c} = \mathcal{A}\hat{a} + \mathcal{B}\hat{a}^\dagger$ and $\hat{c}^\dagger = \mathcal{A}^*\hat{a}^\dagger + \mathcal{B}^*\hat{a}$, where $\mathcal{A}$ and $\mathcal{B}$ are constant factors to be determined. If we are dealing with genuine bosonic operators, a Bogoliubov transformation can be performed such that the commutator satisfies 
\begin{equation}
[\hat{c}, \hat{c}^\dagger] = |\mathcal{A}|^2 - |\mathcal{B}|^2 = 1. 
\end{equation}
This can indeed be achieved by choosing appropriate values for $\mathcal{A}$ and $\mathcal{B}$, as follows:
\begin{equation}\label{OpLad}
    \begin{split}
        \hat{a} &= \frac{1}{2}\sqrt{\frac{\omega_0}{\Omega}}\left[ \left(e^{-i\phi} + 1 \right)\hat{c} +  \left(e^{i\phi} -1 \right)\hat{c}^\dagger\right], \\ 
        \hat{a}^\dagger &= \frac{1}{2}\sqrt{\frac{\omega_0}{\Omega}} \left[\left(e^{-i\phi} - 1 \right)\hat{c} + \left(e^{i\phi} + 1 \right)\hat{c}^\dagger  \right],
    \end{split}
\end{equation}
where $\phi$ is a phase such that $\tan{\phi} = W/\Omega$.  Using (\ref{OpLad}) in (\ref{Haa}) we get,
\begin{eqnarray}\label{BogoH}
\nonumber
        \mathcal{\hat{H}} &=& \left(\frac{\hbar \omega_0^2(e^{-2i\phi} - 1)}{4 \Omega} + \frac{i \hbar W \omega_0 e^{-i\phi}}{2\Omega} \right) \hat{c}^2 \\
        \nonumber
        &+&\left( \frac{\hbar \omega_0^2(e^{2i\phi} - 1)}{4 \Omega} - \frac{i \hbar W \omega_0 e^{i\phi}}{2\Omega}\right) \hat{c}{^\dagger}^2 + \\ 
        &+&\ \left( \frac{\hbar\omega_0^2}{2 \Omega} - \frac{\hbar W \omega_0 \sin{\phi}}{2 \Omega} \right) \left(\hat{c}\hat{c}^\dagger + \hat{c}^\dagger\hat{c} \right).
\end{eqnarray}
The terms proportional to $\hat{c}^2$ and $\hat{c}^{\dagger 2}$ vanish when we explicitly use $\sin{\phi} = W/\omega_0$ and $\cos{\phi} = \Omega/\omega_0$ in Eq.~(\ref{BogoH}). Therefore, the Hamiltonian can finally be written in the form: $\mathcal{\hat{H}} = \hbar \Omega\left(\hat{c}^\dagger\hat{c} + 1/2\right)$.
The quantum statistical-mechanical description of this system is given in the body of the text.


\begin{thebibliography}{9}

\bibitem{dp} M. Laing and M. Laing, J. Chem. Educ. {\bf 83}, 1499 (2006).
\bibitem{MK} C.G. Maier and K. Kelley, J. Am. Chem. Soc. {\bf 54} 3243 (1932).
\bibitem{geology} D. Giordano and J. K. Russell, Che. Geol. {\bf 461}, 96 (2017).
\bibitem{geology2} M. Wu, Z. Liu, Y. Qin, K. Su, and Z. Yu, Rock. Mech. Rock. Eng. {\bf 58}, 8773 (2025).
\bibitem{volc} M. J. Heap, A. R. L. Kushnir, J. Vasseur, F. B. Wadsworth, P. Harl\'e, P. Baud, B. M. Kennedy, V. R. Troll, F. M. Deegan, J. Volcanol. and Geotherm. Res., {\bf 398}, 106901 (2020).
\bibitem{ceramics} D. A. Abreu, A. Schnickmann, T. Schirmer, O. Fabrichnaya, J Am. Ceram. Soc. 108, e20694 (2025).
\bibitem{material} S. S. S{\o}rensen, M. B. {\O}stergaard, M. Stepniewska, H. Johra, Y. Yue, and M. Smedskjaer,  M.ACS Appl. Mater. Interfaces, {\bf 12}, 18893 (2020).
\bibitem{CP-1} E. Boscheto, M. de Souza, A. L\'opez-Castillo, Physica A {\bf 451}, 592 (2016). 
\bibitem{CP0} A. Guha, P. K. Das, Physica A {\bf 495},  18 (2018).
\bibitem{ED0} V. V. Novikov, Journal of Thermal Analysis and Calorimetry {\bf 138}, 265 (2019).
\bibitem{ED1} E. Gamsj|\"ager and M. Wiessner, Entropy {\bf 26}, 452 (2024).
\bibitem{anh} A J E Foreman, Proc. Phys. Soc. {\bf 79} 1124 (1962).
\bibitem{anh2} D. W Field, Aust. J. Phys. {\bf 27} 831 (1974).
\bibitem{anh3} J. M. Keller and D. C. Wallace, Phys. Rev. {\bf 126}, 1275 (1962).
\bibitem{anh4} R. A. Cowley, Rep. Prog. Phys. {\bf 31} 123 (1968).
\bibitem{CP1} W. W. Anderson, AIP Advances {\bf 9}, 075108 (2019).
\bibitem{CP2} T. Cardoso e Bufalo, R. Bufalo and A. Tureanuy, Annals of Phys. {\bf 440}, 168835 (2022).
\bibitem{vassilev} V. P. Vassilieva, A. F. Taldrik, J. Alloys and Compounds {\bf 872}, 15968 (2021).
\bibitem{callen} H. B. Callen, Thermodynamics and an Introduction to Thermostatistics, John Wiley and Soons, New York (1985).
\bibitem{daniel} D. I. Khomskii, Basic aspects of the quantum theory of solids, Cambridge Univ. Press (2010).
\bibitem{pina} C. Pi\~na; E. Mu\~noz; J. L. Bold\'u, J. Chem. Phys. {\bf 79}, 2172 (1983).
\bibitem{gudyma} A. Gudyma and Iu Gudyma, Low Temp. Phys. {\bf 47}, 457?465 (2021).
\bibitem{horbach} A. Bhattacharya, J. Horbach, and S. Karmakar1, Phys. Rev E {\bf 111}, 015429 (2025).
\bibitem{alpha} B. Zhang, X.B.Li, D.Li, Calphad: Computer Coupling of Phase Diagrams and Thermochemistry {\bf 43},  7 (2013).
\bibitem{hassel}  Hassel Ledbetter, Int. J. Thermophys. {\bf 12}, 637, (1991).
\bibitem{chinese} X. Zhang, S. Sun, T. Xu, T. Y. Zhang, Sci China Tech Sci, {\bf 62}, 1565 (2019).
\bibitem{wilson} A. J. C. Wilson, Proc. Phys. Soc. {\bf 53} 235 (1941).
\bibitem{nistCu} https://janaf.nist.gov/tables/Cu-002.html
\bibitem{WM} G. K. White and M. L. Minges,  Int. J. Thermophys., {\bf 18}, 1269 (1997).
\bibitem{brooks} C. R. Brooks and R. E. Bingham, J. Phys. Chem. Solids {\bf 29}, 1553 (1968).
\bibitem{nistAl} https://janaf.nist.gov/tables/Al-002.html
\bibitem{desai} P. D. Desai, Int. J. Thermophys., {\bf 8}, 621 (1987).
\bibitem{khvan} A.V. Khvana, A.T. Dinsdaleb,d, I.A. Uspenskayac, M. Zhilina, T. Babkinac, A.M. Phiria, Calphad {\bf 60} 144 (2018).
\bibitem{young} M. Hasegawa and W. H. Young, J. Phys. F: Met. Phys. {\bf 11}, 977 (1981).
\bibitem{nistPb} https://janaf.nist.gov/tables/Pb-002.html
\bibitem{arblaster} J. W. Arblaster, Calphad, {\bf 39}, 47 (2012).
\bibitem{wei} S. Wei, C. Li, and M. Y. Chou, Phys. Rev. B {\bf 50}, 14587 (1994).
\bibitem{kagaya} H. M. Kagaya, Y. Kitani, and T. Soma, Solid State Comm. {\bf 58}, 399 (1986).
%\bibitem{xu} C. H. Xu, C. Z. Wang, C. T. Chan, and K. M. Ho, Phys. Rev. B {\bf 43}, 5024 (1991).
\bibitem{nistSi} https://janaf.nist.gov/tables/Si-002.html
\bibitem{okhotin} A. S. Okhotin, A. S. Pushkarskij,  V. V. Gorbachev (1972). Thermophysical properties of semiconductors. Atomizdat.
\bibitem{gauster} W. B. Gauster, J. Appl. Phys. {\bf 44}, 1089 (1973).
\bibitem{nistGe} Handbook of Chemistry and Physics, W. M. Heynes (Editor), 95th ed (2015).
\bibitem{yuan} M. Yuan , C. Li, Cuiping Guo , Z. Du, J. Alloys and Compounds {\bf 993}, 174541 (2024).
\bibitem{cupid} T. L. Reichmann, D. Li, and D. M. Cupid, Phys. Chem. Chem. Phys {\bf 20}, 22856 (2018).
\bibitem{denisova0} L. T. Denisova,  L.A. Irtyugo, Y.F. Kargin, N. V. Belousova, V. V. Beletskii, and V. M. Denisov, Inorg. Mater. {\bf 54}, 361 (2018).
\bibitem{robie} R. A. Robie, B. S. Hemihgway, and J. M. Fisher, Thermodynamic Properties of Minerals and Related Substances at 298.15 K and 1 Bar (10$^5$ Pascal) and at higher Temperatures. United States Government Printing Office: Washington (1979).
\bibitem{mostafa} A. T. M. G. Mostafa, J. M. Eakman, M. M. Montoya, and S. L. Yarbro, Ind. Eng. Chem. Res. {\bf 35}, 343 (1996).
\bibitem{denisova} L.T. Denisova,  D.V. Belokopytova, Y. F. Kargin, G. V. Vasil?ev, N. V. Belousova, and V. M. Denisov. Russ. J. Inorg. Chem. {\bf 69}, 1352 (2024).
\bibitem{matthias} M. T. Agne, K. Imasato, S. Anand, K. Lee, S. K. Bux, A. Zevalkink, A. J.E. Rettie, D. Y. Chung, M. G. Kanatzidis, G. J. Snyder, Mater. Today Phys., {\bf 6}, 83 (2018).


\end{thebibliography}
\end{document}